\documentclass[10pt,a4paper,twocolumn]{article}
\usepackage[dvips]{graphicx}

\let\oldcaption\caption
\renewcommand{\caption}[2][1]{\oldcaption{{\small \sl #2}}}

\begin{document}

 \begin{center}
   {\LARGE Improved Surface Treatment \\ 
    of the Superconducting \\ 
    TESLA Cavities\footnote{Article published in Nucl. Inst. Meth. A
    516 (2-3) (2004) pp.213-227.}\\}
    \vspace{0.8cm}

    L. Lilje \footnote{Corresponding author: Lutz.Lilje@desy.de}, A. Matheisen, D. Proch, \\
    D. Reschke, D. Trines,
    
    {\it DESY, Notkestrasse 85, D-22607 Hamburg, Germany}
    \vspace{0.3cm}
    
    C. Antoine, J.-P. Charrier, H. Safa, B. Visentin,
    
    {\it CEA Saclay, DAPHNIA, 91191 Gif-sur-Yvette, France}
    \vspace{0.3cm}
    
    C. Benvenuti, D. Bloess, E. Chiaveri, L. Ferreira, R. Losito, H.
    Preis, H. Wenninger
    
    {\it CERN, CH-1211 Geneva 23, Switzerland}
    \vspace{0.3cm}
    
    P. Schm\"user
    
    {\it Universit\"at Hamburg, Notkestrasse 85, D-22607 Hamburg,
      Germany}
\end{center}

{\bf Abstract}

The proposed linear electron-positron collider TESLA is based on 1.3
GHz superconducting niobium cavities for particle acceleration. For a
center-of-mass energy of 500 GeV an accelerating field of 23.4 MV/m is
required which is reliably achieved with a niobium surface preparation
by chemical etching.  An upgrade of the collider to 800 GeV requires
an improved cavity preparation technique.  In this paper results are
presented on single-cell cavities which demonstrate that fields of up
to 40 MV/m are accessible by electrolytic polishing of the inner
surface of the cavity.



\section{Introduction}
Linear electron-positron colliders for physics in the Higgs and
supersymmetric particle regime have to reach centre-of-mass energies
in excess of 500 GeV, and hence high accelerating fields in
the accelerating structures are needed to limit the size of the
accelerator complex. TESLA \cite{tesla_cavity_tdr} is
the only collider project based on superconducting cavities.  In the
500 GeV baseline design (TESLA-500) the accelerating field, often
called ''gradient'' hereafter, amounts
to 23.4 MV/m. The 1.3 GHz nine-cell niobium cavities for the TESLA Test
Facility (TTF) linac have achieved an
average gradient of 26.1 $\pm$ 2.3~MV/m at a quality factor $Q_0 \ge 1
\cdot 10^{10}$ in the most recent industrial production of 24
cavities. The technology developed for TTF is hence adequate for
TESLA-500 but considerable improvements are needed for an upgrade of
the collider to 800 GeV (TESLA-800). A detailed description of the
present status of the nine-cell cavity layout, fabrication,
preparation and tests can be found in \cite{tesla_cavities}.

\begin{figure}[t!]
  \begin{center}
     \includegraphics[width=8cm]{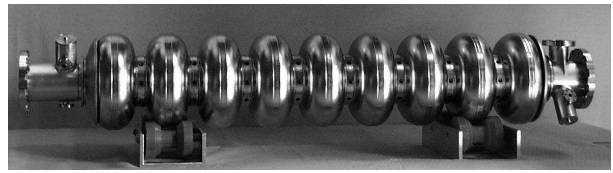}
  \end{center}
    \caption{Superconducting 1.3 GHz 9-cell cavity for the TESLA Test
      Facility.}
    \label{fg:cavity}
\end{figure}

The TESLA cavities (Figure \ref{fg:cavity}) are made from 2.8 mm thick
niobium sheets by deep drawing and electron beam welding. A damage
layer of about 100 $\mu$m thickness has to be removed from the inner
surface to obtain an optimum performance in the superconducting
state. For the TTF cavities this has been done so far by chemical
etching, see below. Using an alternative preparation method,
electrolytic polishing (or electropolishing for short), scientists at
the KEK laboratory in Tsukuba (Japan) have achieved gradients of up to
40 MV/m in 1.3~GHz single-cell cavities \cite{saito_superiority}. This
was the motivation for starting an R\&D program on electropolishing of
single-cell cavities of the TESLA shape. This program has been carried
out in collaboration between CERN, DESY and Saclay.

\section{Niobium Chemistry}
\label{sec:chemistry_basics}
Niobium metal has a natural Nb$_2$O$_5$-layer with a thickness of
about 5~nm. Below this layer other oxides and sub-oxides can be found
\cite{grundner_77,grundner_80,dacca_98}.  Nb$_2$O$_5$ is chemically
rather inert and can be dissolved only with hydrofluoric acid
(HF). The strong acids used in the niobium surface treatment
like HF, HNO$_3$ or H$_2$SO$_4$
play also an essential role for removing defects from the inner
surface of the cavity either by dissolving the foreign material itself
or the surrounding niobium. Such defects may be abrasive particles from
grinding, imprints from the deep drawing process, niobium protrusions
from scratches or dirt particles sticking to the surface. Cleaning the
niobium with chemical methods is the most practical way to achieve a
high quality superconducting surface.  Chemical or electrolytic
processes can be applied to niobium resonators if the safety 
and environmental standards regarding HF-containing acid mixtures are
obeyed. Nitrous gases, oxygen and hydrogen are produced during etching
or electropolishing.  A review of the preparation methods of niobium
cavities is given in \cite{kneisel_kfk}.

\subsection{Chemical etching}
\label{sec:etching}
The sheet rolling of niobium produces a damage layer of about 100
$\mu$m thickness which has to be removed in order to obtain a surface
with excellent superconducting properties. The standard procedure is
chemical etching which consists of two basic steps: dissolution of the
Nb$_2$O$_5$ layer by HF and re-oxidation of the niobium by a strongly
oxidizing acid \cite{gmelin_nb,siemens_scrf}. Then the new oxide layer
will be dissolved by the HF again. The most frequently used oxidizing
agent for niobium is nitric acid (HNO$_3$). The material removal rate
in a mixture of HF~(40\%) and HNO$_3$~(65\%) is large, of the order of
30~$\mu$m per minute \cite{siemens_scrf}.  The reactions are strongly
exothermic and can cause a thermal runaway situation. Additionally,
large quantities of gases (hydrogen, nitrous gases, HF) 
are produced. For these reasons the mixture of HF and HNO$_3$ alone is
not suitable for the etching of cavities.

To obtain a better process control a buffer substance like phosphoric
acid H$_3$PO$_4$ (concentration of 85\%) may be added
\cite{guerin_bcp2}. Furthermore the mixture is cooled below
15$^\circ$C which also reduces the migration of hydrogen into the
niobium lattice \cite{siemens_scrf,roeth}. The standard procedure with
a removal rate of about 1~$\mu$m per minute is called {\it buffered
chemical polishing}\footnote{The term ``polishing'' is not really
adequate since the resulting surface roughness is in the $\mu$m range
and grain boundaries are enhanced. }  (BCP) with an acid mixture
containing 1~part~HF, 1~part~HNO$_3$ and 2~parts~H$_3$PO$_4$ in
volume. Another possibility is to use concentrated sulphuric acid
(H$_2$SO$_4$) as a buffer \cite{siemens_scrf,claire_99}.  At TTF, a
closed-circuit chemistry system is used in which the acid is pumped
from a storage tank through a cooling system and a filter into the
cavity and then back to the storage. The gases produced are not
released into the environment without prior neutralization and
cleaning. The cavities are rinsed with low pressure ultrapure water
immediately after the chemical treatment.

\begin{figure}[!ht]
  \centering
    \includegraphics[angle=-90,width=7cm]{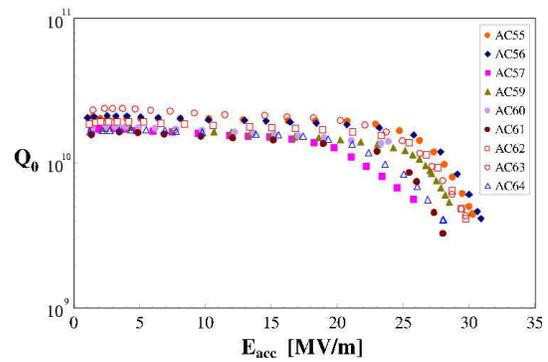}
  \caption{Excitation curves of several nine-cell TESLA cavities  with surface
    preparation by chemical etching (BCP). The quality factor is
    plotted as a function of the accelerating field. Test temperature
    is 2~K.}
  \label{fig:ninecell_prod}
\end{figure}

By now there exists compelling evidence that the BCP process limits
the attainable accelerating fields of multi-cell cavities to about
30~MV/m even if niobium of excellent thermal conductivity is
used. Typical accelerating gradients achieved with BCP in defect-free
cavities are shown in figure \ref{fig:ninecell_prod}. Measurements on
etched one-cell cavities in other laboratories yielded similar results
\cite{kneisel_97,safa_97,saito_superiority}. Etched cavities exceeded
30 MV/m only in rare cases \cite{kneisel_high_field,visentin_02}.

The etching process is accompanied with undesirable effects such as
migration of hydrogen into the bulk niobium and strong grain boundary
etching. The first effect can be reduced by cooling the acid below
15$^\circ$C \cite{roeth}. In principle, grain boundary etching could
be suppressed by using an acid mixture with a high etching rate. This,
however, would not be advisable in the nine-cell TESLA cavity with an
inner surface of about 1~m$^2$ since the large amount of produced heat
would speed up the reaction even more and preclude a well-controlled
etching process.

\subsection{Electrolytic polishing}
\label{sec:electropolishing}
An alternative method to etching is electrolytic polishing or {\it
electropolishing} (EP) in which the material is removed in an acid
mixture under the flow of an electric current. Sharp edges and burrs
are smoothed out and a very glossy surface can be obtained.  The
electric field is high at protrusions so these will be dissolved
first. On the other hand, the field is low in the grain boundaries and
little material will be removed here while in the BCP process strong
etching is observed in the boundaries between grains.

\begin{figure}[tbp]
    \centering
    \includegraphics[height=4cm, width=7cm]{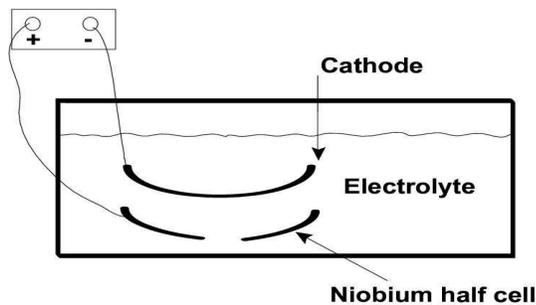}

    a) 

    \includegraphics[height=4cm, width=7cm]{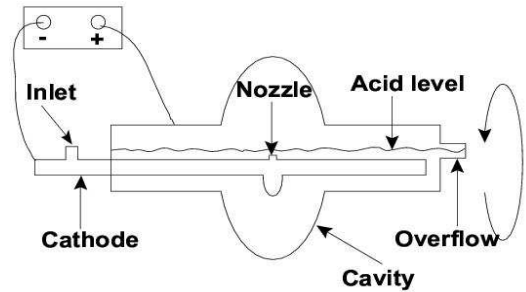}

    b)
\caption{ a) Schematic of a half-cell EP system.  b: Schematic of a
    cavity EP system. Detailed descriptions  of the two systems are
    given in the text.}
  \label{fig:ep_scheme}
\end{figure}

Electropolishing of niobium cavities has been known for 30~years.
The most widely used electrolyte is a mixture of concentrated HF and
concentrated H$_2$SO$_4$ in volume ratio of~1:9~\cite{siemens}. A
pulsed electric current was used in a horizontal EP setup for
superconducting niobium cavities prepared at CERN in collaboration with
Karlsruhe in 1979 \cite{kfk_ep_hor_79}. A continuous method for
horizontal EP has been developed at KEK in 1989 \cite{saito_ep_system}.

The chemical processes are   as follows
\cite{kneisel_kfk,ponto}:
\begin{eqnarray}
  2 \textrm{Nb} + 5 \textrm{SO}_4^{--} + 5 \textrm{H}_2\textrm{O}
  & \to & \textrm{Nb}_2\textrm{O}_5 + 10 \textrm{H}^+ \nonumber\\
  &     &  + 5\textrm{SO}_4^{--}  + 10 \textrm{e}^- \nonumber \\
  \textrm{Nb}_2\textrm{O}_5 +6 \textrm{HF} & \to &
  \textrm{H}_2\textrm{NbOF}_5 \nonumber \\ 
  &     &+ \textrm{NbO}_2\textrm{F} \cdot 0.5
  \textrm{H}_2\textrm{O} + 1.5 \textrm{H}_2\textrm{O} \nonumber  \\
  \textrm{NbO}_2\textrm{F} \cdot 0.5 \textrm{H}_2\textrm{O} + 4 \textrm{HF} &
  \to & \textrm{H}_2\textrm{NbF}_5 + 1.5 \textrm{H}_2\textrm{O} \nonumber
\end{eqnarray}

Most systems for the EP of accelerator cavities are horizontal as
shown in figure \ref{fig:ep_scheme}. The advantage is that the
produced gases (mainly hydrogen) are rapidly removed from the wetted
niobium surface \cite{kfk_ep_hor_79}.  Gas bubbles sticking on the
surface could lead to etching pits. In a vertical setup these bubbles
would move slowly upwards and create axial wells. A
drawback of the horizontal arrangement is that the cavity has to be
rotated. There is some difficulty to achieve a leak tight rotary sleeve
for the acid mixture. In addition, the removal rate is reduced by a
factor of two since the surface is immersed only half of the time in
the acid to allow the hydrogen gas to escape through the upper part of
the beam tube.

\subsection{Comparison of etched and electropolished surfaces}
\label{sec:comp-etch-electr}

\begin{figure}[htbp]
  \begin{center}
      \includegraphics[height=6cm, width=7cm]{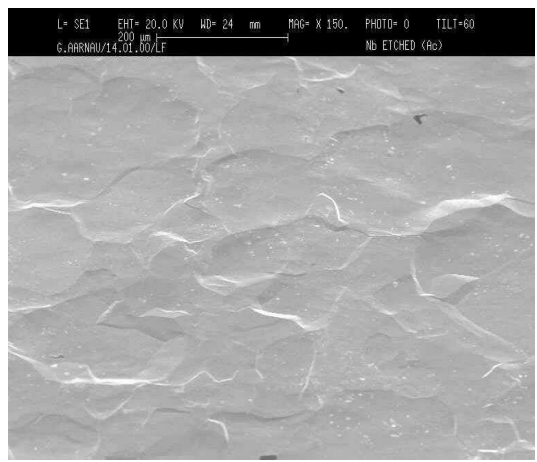}

      a) $400 \times 800 \mu$m

      \includegraphics[height=6cm, width=7cm]{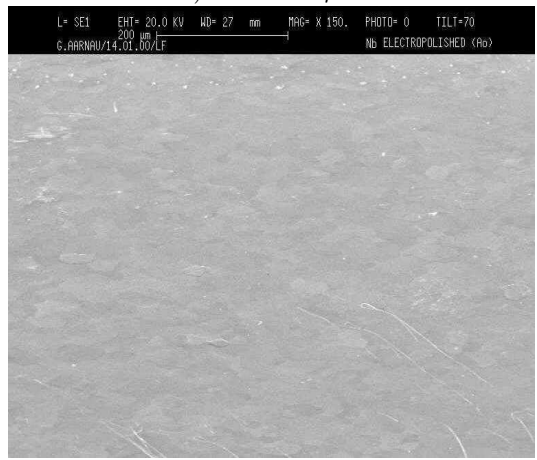}

      b) $400 \times 800 \mu$m

    \caption[Niobium surfaces after
    etching and electropolishing.]{\label{fig:niobium_surfaces}
      Niobium surfaces after etching (a) and electropolishing
      (b). SEM micro-graphs are courtesy of G. Arnau, CERN.}
  \end{center}
\end{figure}

\begin{figure}[htbp]
  \begin{center}
   \includegraphics[angle=90, width=7cm]{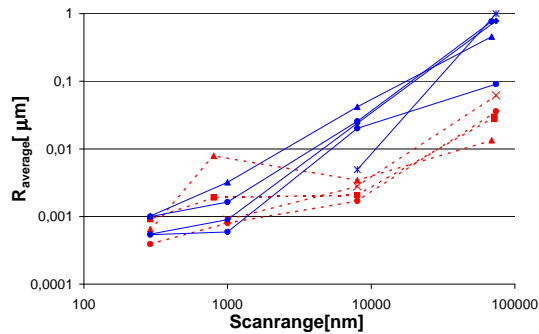}
    \caption[Average roughness as a function of the scan length
    of the atomic force microscope.]{Average roughness as a function
      of the scan length of an atomic force microscope. EP treated
      samples are shown with dotted 
      lines, while BCP samples are shown with full lines.  At large
      scan length EP yields about one order of magnitude lower average
      roughness than BCP while for a short scan length (less than the
      grain size) the difference  becomes smaller.  }
    \label{fig:roughness_samples}
  \end{center}
\end{figure}

Micro-graphs of  BCP and EP treated niobium
samples are compared in figure \ref{fig:niobium_surfaces}.
One can see that EP smoothes out the grain
boundaries far better than BCP. The average roughness of chemically
etched niobium surfaces is of the order of $1\mu$m
\cite{claire_99} while EP surfaces are at least one
order of magnitude smoother, see figure \ref{fig:roughness_samples}.
It should be noted, however, that the measured unevenness of the
surface depends on the scan length of the atomic force microscope.  On
a nanometer scale the roughness of BCP and EP treated samples is
comparable.  This can be understood because at such small lengths the
measurement takes place on a single niobium grain. The main difference
between EP and BCP is the smoothening of the ridges at grain
boundaries which is measured only when the scan length is sufficiently
large.


On etched surfaces the step height at grain boundaries can be a few
$\mu$m.  Particularly large steps may occur at the electron-beam welds
at the cavity equator. For non-optimized weld parameters steps as high
as 30~$\mu$m have been observed \cite{geng_99}.  The steps at grain
boundaries and in the weld seams may lead to a magnetic field
enhancement and a premature breakdown of superconductivity
\cite{liepe_99}.
This interpretation is corroborated by the observation that several
chemically etched cavities exhibited a quench in the region of high
magnetic field  near the equator weld without any indication of
contamination by foreign materials. It was possible to shift the
quench position along the equator region by applying a new chemical
etching, but the maximum achievable field was not improved
\cite{kako_97,kako_99}. 

An important question for electropolishing is of course how much
material has to be removed to achieve a smooth surface.  On
electropolished samples the  roughness drops below 1~$\mu$m after
150~$\mu$m of material have been removed.
This corresponds to the thickness of the damage layer which has to be
removed anyway.  Therefore an electropolishing of at least 100~$\mu$m
is a reasonable choice both for  surface smoothening and damage layer
removal. 

\subsection{Description of the  EP systems}
\subsubsection{Half-cell electropolishing}
\label{sec:halfcell_EP}
In the framework of the DESY-CERN-Saclay
 collaboration a setup for the EP of half-cells
was developed at CERN. The goal was to investigate the viability of a novel
cavity fabrication and preparation concept: removal of most of the
damage layer by EP on simple subunits and assembly of the multi-cell
cavity from pre-polished half-cells.  An electrolyte developed at CERN
\cite{guerin_74,guerin_82} was used (see table
\ref{tab:ep_parameters}) with a removal rate of 1~-~2~$\mu$m per
minute. The electrode was made from copper.  An electropolishing of
about 100 $\mu$m was applied to the half cells.  Then 4~one-cell
cavities were electron-beam welded at an industrial company with a
standard electron-beam welding apparatus (vacuum $10^{-5}$ mbar), and
11 cavities were welded at CERN under much better vacuum conditions
(pressure in the welding chamber below $ 10^{-6}$mbar).

Two of the cavities were tested immediately after welding and
high-pressure water rinsing. Not unexpectedly, the results were poor:
the cavities reached a maximum accelerating field of just 8 - 10 MV/m
with a strong decrease in quality factor. A slight chemical etching
which would have been sufficient to dissolve a possible copper
deposition on the niobium surface cured the Q-degradation, however electron
field emission was still present at low excitation implying that
another contaminant besides copper was present. After a chemical
etching (BCP) of 20~-~50~$\mu$m both cavities reached the standard
performance (25~MV/m) of BCP-treated cavities. The obvious explanation
is that welding deposits  on the inner surface of the cavity must be
removed by a BCP or EP of 20~-~50~$\mu$m to arrive at a defect-free
surface.  

The conclusion to be drawn from these observations is that the
fabrication of a multi-cell cavity from prepolished half-cells is
indeed a viable concept which would substantially reduce the amount of
EP needed in the finished resonator.

\begin{table}[btp]
  \centering
  \begin{tabular}[t]{|p{4.2cm}|p{3.3cm}|}
\hline
{\bf EP for half cells}    & 24 \% ~ HF~(40\%)          \\
Acid mixture               & 21 \% ~ H$_2$SO$_4$~(96\%) \\
                           & 38 \% ~ H$_3$PO$_4$~(83\%) \\
                           & 17 \% ~ Butanol     \\
\hline
Voltage                    & 4 - 6       ~ V \\
Current density            & 1 - 1.4   ~ A/cm$^2$ \\
Removal rate               & 1 - 2       ~ $\mu$m/min\\
\hline
\hline
{\bf EP for cavities}      & 10 \% ~ HF~(40\%) \\
Acid mixture               & 90 \% ~ H$_2$SO$_4$~(96\%) \\
\hline
Voltage                    & 15 -20      ~ V \\
Current density            & 0.5 -0.6      ~ A/cm$^2$ \\
Removal rate               & 0.5         ~ $\mu$m/min\\
Temperature of electrolyte & 30 - 35     ~ $^\circ$C \\
Rotation                   & 1           ~ rpm \\
Acid flow                  & 5           ~ liters/s \\
\hline
  \end{tabular}
  \caption{Parameters for the EP of half-cells and the EP of
    one-cell cavities at CERN.}
  \label{tab:ep_parameters}
\end{table}

\subsection{Electropolishing of complete cavities}
\label{sec:cavity_ep}
A setup for the electropolishing of single-cell cavities was built at
CERN.  The parameters were chosen similar to those of the successful
KEK system \cite{saito_ep_system} and are summarized in table
\ref{tab:ep_parameters}.  The essential difference is the cathode
material: pure aluminum at KEK and copper at CERN. Both materials
have their respective advantages and disadvantages. The copper
electrode requires some care, see below. The cathode is surrounded
with tube made from a porous PTFE\footnote{Polytetrafluoroethylene,
  for example Teflon\textregistered} cloth to keep the
electrolytically produced hydrogen gas away from the niobium surface.

Except for the cathode all components of the EP system are made from
plastic material which is inert against the aggressive acid mixture.
All parts in contact with the HF should be made from fluoroplastics
such as PFA\footnote{Polyperfluoroalkoxyethylene},
PVDF\footnote{Polyvinylidene Fluoride} or
PTFE.  The EP system is placed in a vented area
where the exhaust gases are pumped through a neutralization system to
avoid environmental hazards.  The acid mixture is contained in a
closed-circuit.  The electrolyte is stored in containers cladded with
Teflon which can be cooled by water flowing through Teflon-covered
piping. The volume above the electrolyte is filled  with dry
nitrogen to avoid water vapour absorption by the strongly hygroscopic
H$_2$SO$_4$. The acid mixture is pumped with a membrane pump through a
cooler and a filter with 1 $\mu$m pore size before it reaches the
inlet of the cathode. Then the electrolyte flows through the cathode
to the center of the cell (see figure \ref{fig:ep_scheme}(b)). The
acid returns to the storage tank via an overflow.

The cavity is installed horizontally together with the cathode and
then the lower half is filled with the electrolyte which attacks the
niobium only very slowly when no voltage is applied (etch rate less
than 1 nm per hour).  After the equilibrium filling level has been
reached, the rotation is switched on and a leak check is done. Then
the current is switched on and the current-voltage characteristic is
measured. At a certain voltage (see table \ref{tab:ep_parameters})
current oscillations set in which are an indication that two
alternating processes take place: dissolution of Nb$_2$O$_5$ by HF and 
re-oxidation by H$_2$SO$_4$.  The best polishing results are obtained
for an oscillation amplitude of 10~-~15\% around the mean value.
The temperature of the acid has to be around 30~-~35$^\circ$C during
the EP. Temperatures above $40^\circ$C must be avoided as they result
in etching pits on the surface.

When the desired amount of material has been removed, the current is
switched off. The rotation is stopped and the cavity is put into
vertical position to release the acid mixture. After rinsing with pure
water the cavity and the electrode are dismounted. Another
low-pressure water rinsing follows and the wet cavity is taken into a
glove-box with nitrogen atmosphere. Here the cavity is rinsed with an
ultrapure high-pressure water jet to remove remaining chemical
residues. A rinsing with filtered ethanol follows to speed up the
drying process. The cavity is then stored overnight in a vacuum of
$10^{-3}$~mbar. Then the cavity is either rinsed
with high-pressure water for the performance tests at CERN or filled
with nitrogen gas and sent to Saclay or DESY for high-pressure water
rinsing and tests.

It should be mentioned that the copper electrode used in the CERN
system has a disadvantage.  While the current is flowing, the copper
becomes passivated but after the current has been switched off copper ions
can be dissolved and precipitate on the cavity surface. It has been
observed in a test that this leads to an increased residual resistance
and to strong field emission at low field ($< 10~$MV/m)
\cite{lilje}. To remove the copper deposition 
the cavities are rinsed in sequence
with HNO$_3$~(60 \%), water, HF and water again. These steps are
repeated twice.  The HNO$_3$ dissolves the copper and oxidises the
niobium.  Hydrofluoric acid removes the oxide layer and several
contaminants as indicated by our surface studies.
The overall material removal by this procedure is only
about 10~nm and does not deteriorate the smoothness of the
electropolished surface.

\section[Measurements on  electropolished cavities]
{Measurements on electropolished cavities}
\label{sec:ep_measurements}
\subsection{First tests}
\begin{figure}[!ht]
  \begin{center}
      \includegraphics[angle=90, width=7.5cm]{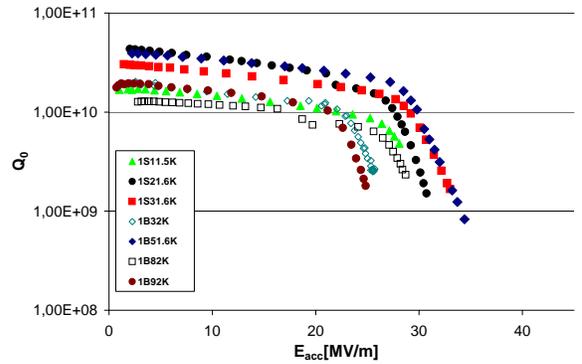} \newline 

      a)

      \includegraphics[width=7.5cm]{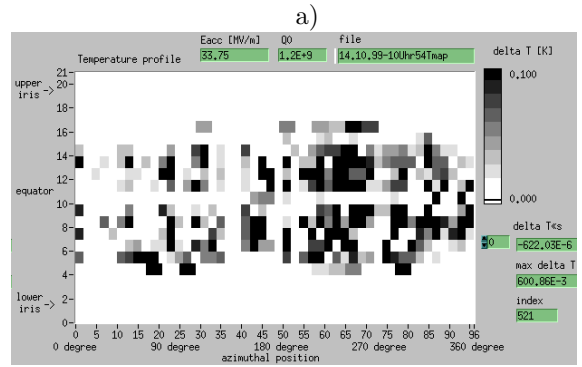} \newline 

      b)
    \caption{a) First tests  of electropolished single-cell
      resonators. Note the strong degradation of  the quality factor
      at accelerating fields $E_{acc}=25~-~30~$ MV/m.
      b) Temperature map of an electropolished single-cell
      resonator.  Shown is an unwrapped view of the outer cavity
    surface. The heating of the surface (dark spots) takes place
      near the equator in the region high magnetic field. }
    \label{fig:ep_no_bake}
  \end{center}
\end{figure}

In their first tests the cavities which were electropolished in the
CERN system showed an unexpected performance limitation: the excitation
curves exhibited a strong degradation in quality factor at high field
as can be seen in figure \ref{fig:ep_no_bake}. Field emission of
electrons could be excluded as an explanation for the performance degradation
since neither X rays nor secondary electrons were observed.
Temperature mapping revealed a global heating in the areas of high
magnetic fields around the equator indicating a rapid increase of the
microwave surface resistance towards high microwave fields.

\begin{figure}[htbp]
  \centering
    \includegraphics[width=7cm]{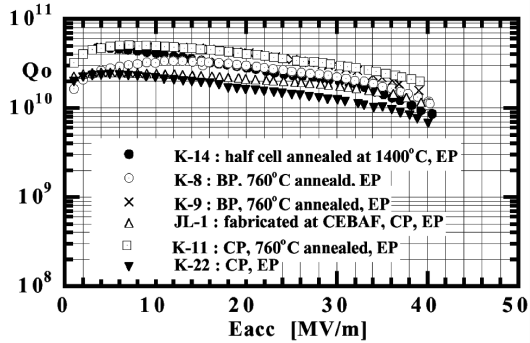}
  \caption[Results from electropolished single cell resonators at
    KEK.]{Results from electropolished single cell resonators at KEK
    \cite{kako_99}. No evidence of a strong degradation of the quality
    factor is seen. As part of the standard
    preparation at KEK  a low-temperature baking at
    85~-~120$^{\circ}$C was applied to the evacuated cavities. 
    Test temperature was 1.5~-~1.7~K. (Courtesy K. Saito, E. Kako)}
  \label{fig:kek_ep_results}
\end{figure}

In contrast to this observation, the excitation curves of the
EP-treated cavities at KEK showed only a moderate drop of the quality
factor and reached much higher accelerating fields, see figure
\ref{fig:kek_ep_results} \cite{saito_high_field,saito_superiority,kako_99}.

\begin{figure}[htbp]
  \begin{center}
      \includegraphics[angle=90, width=7cm]{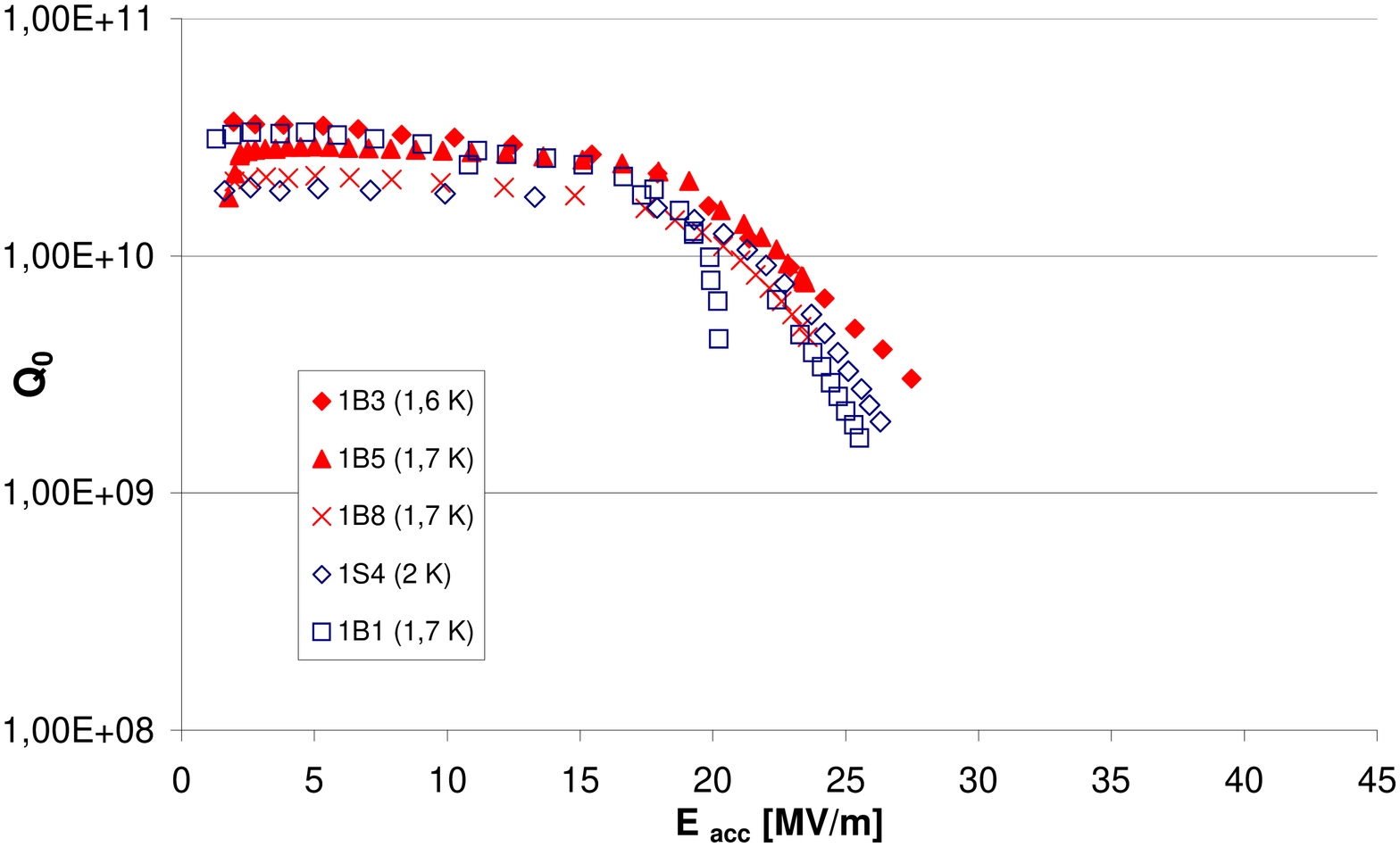} \newline 
      a)

      \includegraphics[angle=0,width=7cm]{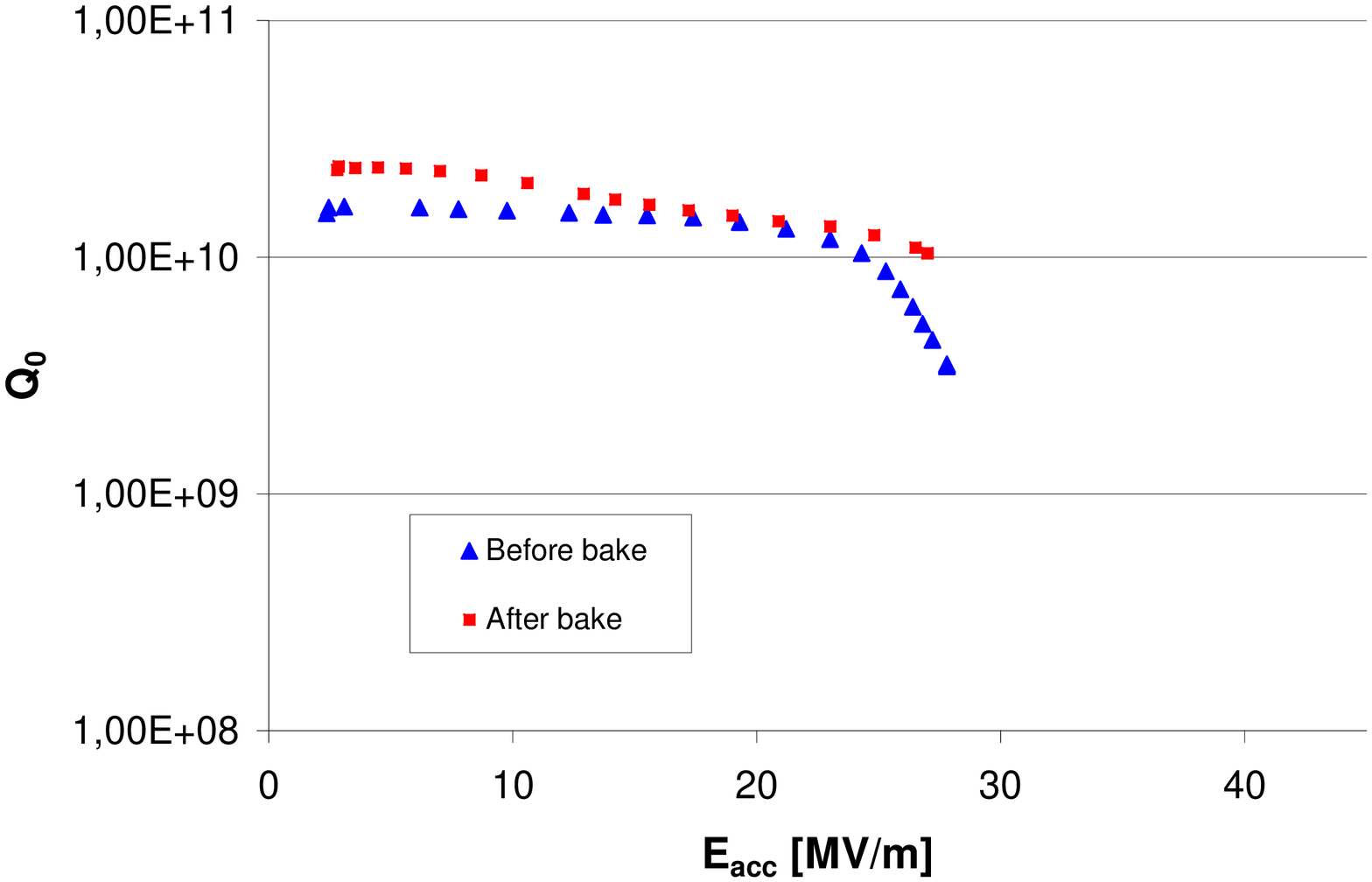} \newline 
      b)
    \caption{a) Excitation curves of etched single cell
      resonators showing a strong degradation of the quality factor
      at $E_{acc}$~=~25~MV/m.   b) Influence of the bakeout
    procedure on the quality   factor $Q_0$  of a BCP-treated cavity.
    Test temperature was 2~K.}
    \label{fig:bcp_no_bake}
  \end{center}
\end{figure}

At Saclay a similar strong $Q$ degradation was observed for chemically
etched cavities \cite{kako_97,safa_97,visentin_bake_98}, see figure
\ref{fig:bcp_no_bake}. The temperature mapping revealed a global
heating of the surface similar to that in figure \ref{fig:ep_no_bake}b.
It was discovered  that the unsatisfactory performance could be
considerably improved by applying a moderate thermal treatment to the
finished cavity \cite{visentin_bake_98}. This
will be called {\it bakeout} (or {\it in situ bakeout}) hereafter.
The procedure at Saclay was as follows. After the last 
high-pressure water rinsing the cavities were evacuated and then
heated up to 170$^\circ$C for 70 hours.  The remarkable observation was
that this low-temperature baking improved the quality factor at the
highest field by nearly a factor of 3.  This result was confirmed in
other tests on BCP-treated cavities. However,  tests revealed also
that the baking procedure did not increase the maximum field level in
chemically etched cavities \cite{visentin_bake_98}.

\subsection{Application of low-temperature bakeout to EP cavities}

\begin{figure}
\begin{center}
\includegraphics[angle=90,width=7cm]{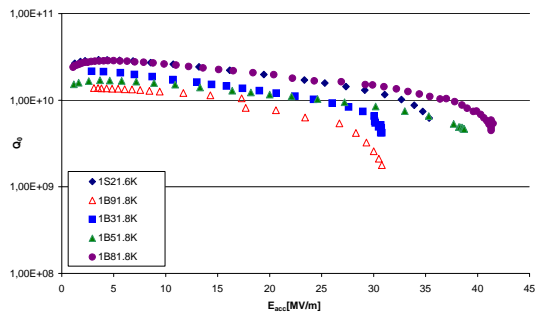}
\caption{\label{fig:ep_onecell} Excitation curves of
  electropolished one-cell cavities after the low-temperature
  bakeout. The tests   have been performed under
    slightly different conditions (different magnetic shielding in
  test cryostats and temperature).}
\end{center}
\end{figure}

Building on the experience with BCP cavities at Saclay and with EP
cavities at KEK (where the bakeout at 85--100$^\circ$ C had been part
of the standard preparation) it was decided to apply the bakeout to
the EP cavities of the CERN-DESY-Saclay collaboration. Figure
\ref{fig:ep_onecell} shows that a dramatic improvement in performance
is achieved: several single cell cavities reach now accelerating
gradients of up to 40 MV/m with quality factors above $5\cdot
10^9$. This behaviour is very similar to the observations made at KEK.

The heat treatment of niobium cavities is a well-known method to
improve the performance. The temperature is in general quite high
though: the TTF cavities are heat treated in an UHV furnace at
800$^\circ$C for hydrogen degassing and stress annealing and
afterwards at 1400$^\circ$C for increasing the thermal conductivity of
the bulk niobium\footnote{During this furnace treatment a titanium
  getter layer is evaporated onto the niobium surface which
protects the niobium from being oxidized by the residual oxygen in
the furnace atmosphere.}. The surprising observation made with the bakeout
effect is that a thermal treatment at such a low temperature, where
the diffusion of any gases dissolved in the niobium lattice is
extremely slow, has a major influence on the high-gradient
performance.  It it obvious that only the thin surface layer
which is essential for the microwave superconductivity can be modified
by the bakeout.

\subsection{High-field performance at different helium
  temperatures}
\begin{figure}[!htbp]
  \begin{center}
    \includegraphics[angle=0,width=7cm]{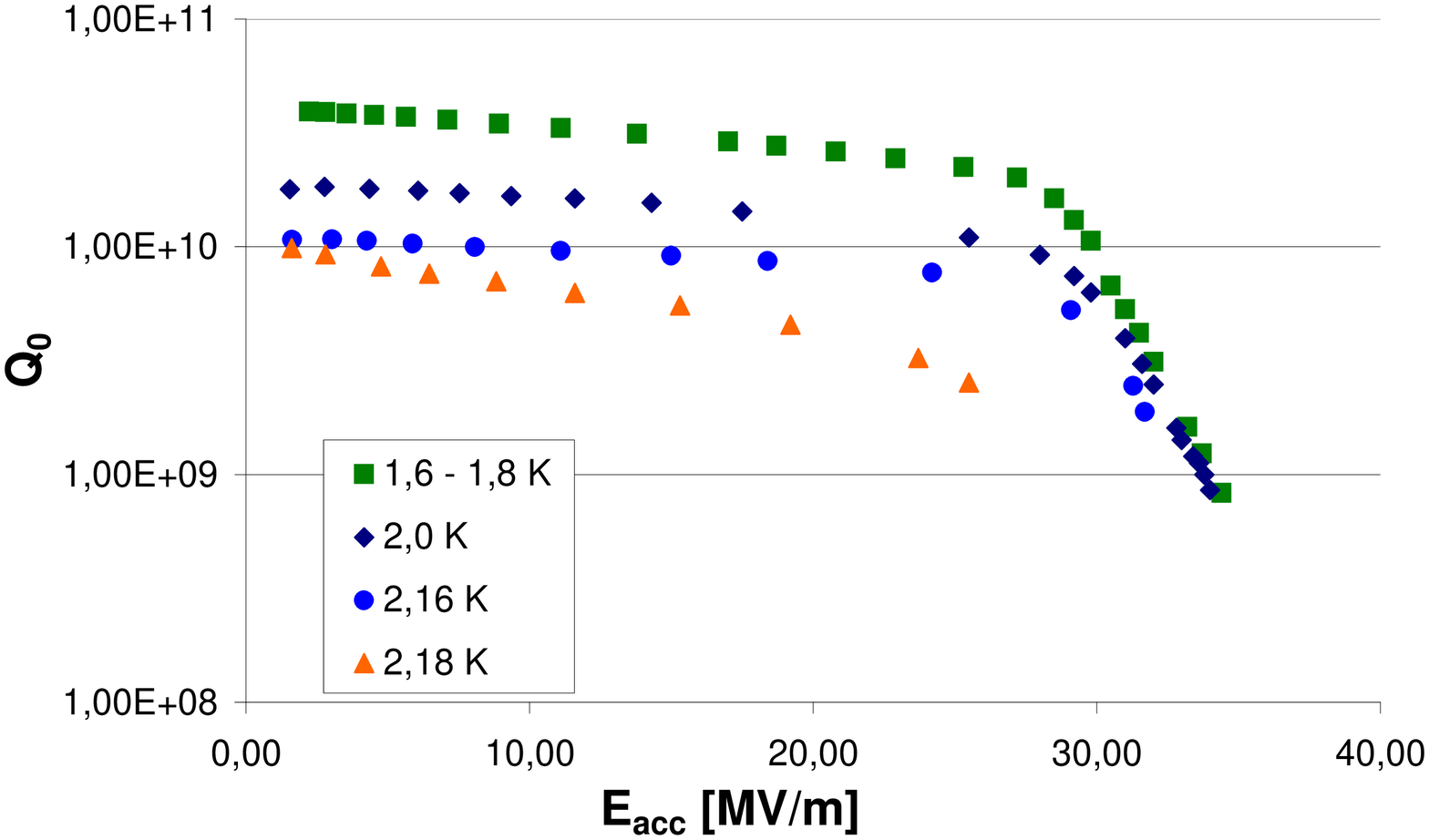} \newline a)

    \includegraphics[angle=-90,width=7cm]{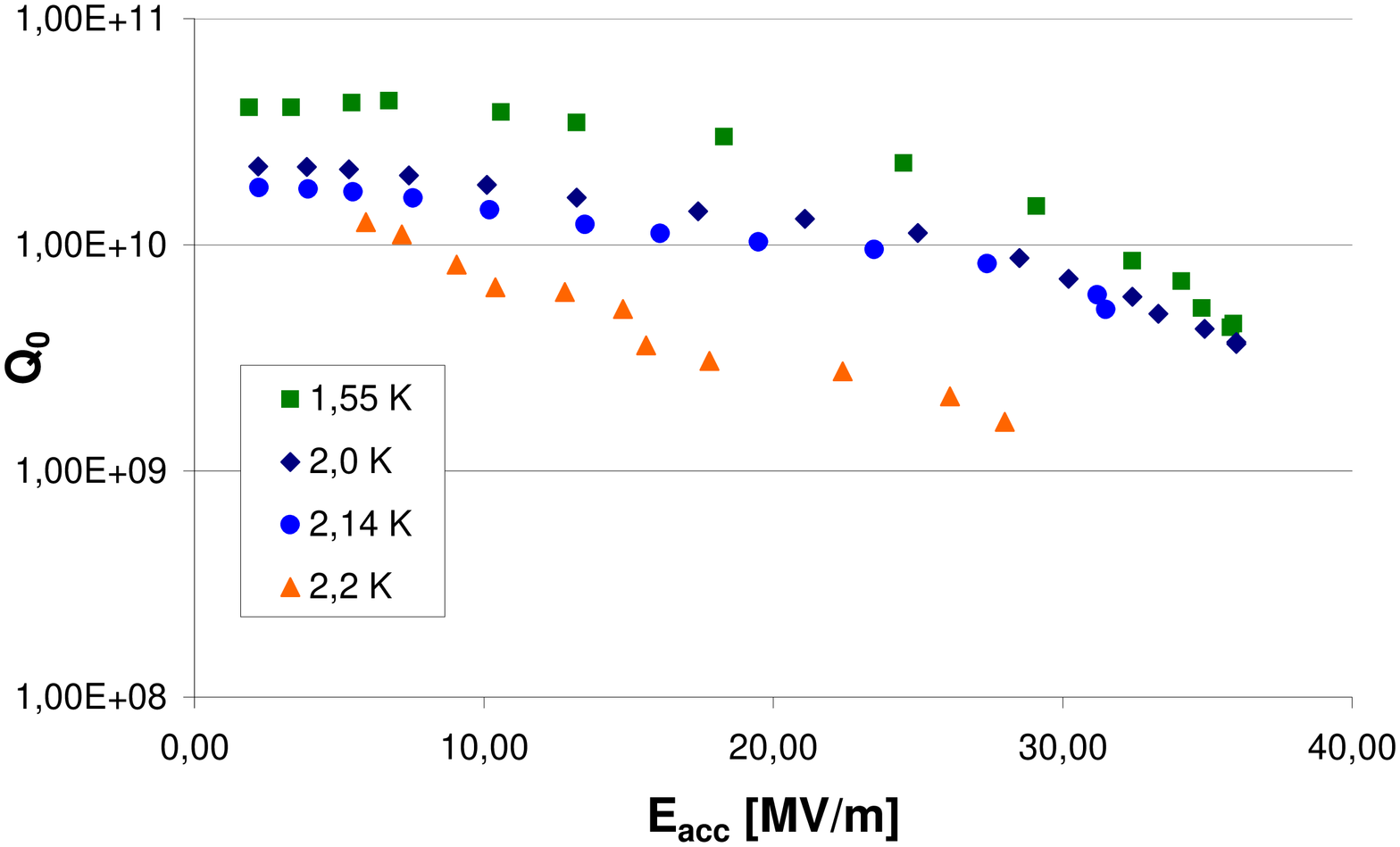} \newline b)
  \end{center}
\caption{Excitation curves of an electropolished cavity
    before (a)) and after  bakeout (b)) for different liquid
    helium temperatures.  }
  \label{fig:bake_qe_temp}
\end{figure}

Figure \ref{fig:bake_qe_temp} shows the excitation curves of an
electropolished cavity before and after the bakeout at helium bath
temperatures between 1.5~K and 2.2~K. At low field the quality factor
exhibits the well-known temperature dependence which is caused by the
exponential temperature dependence of the BCS surface resistance (see below).
However, the maximum achieved gradient and the corresponding $Q_0$
value are almost independent of the bath temperature as long as the
helium coolant is in the superfluid state.  These results are
consistent with thermal model calculations \cite{reschke_97}.  Only
when the temperature is in the vicinity of the Lambda-point (2.17~K) of
liquid helium, where the transition from the superfluid to the
normal-fluid phase takes place, a degradation of the quality factor due
to insufficient cooling can already be seen at low field.

\subsection{High temperature heat treatments of {electropolished}
 cavities}
\label{sec:high-temp-heat}
\label{sec:q_disease}
Electrolytic polishing generates hydrogen which can penetrate into
the niobium lattice.  If too much hydrogen is dissolved 
the danger exists that niobium hydride compounds are formed if
the material is exposed to temperatures around 100~K for an extended
period. These Nb-H compounds have very high microwave losses and may
lead to a reduction of the quality factor by one or two orders of
magnitude. This unfortunate effect has been named {\it Q-disease}.

The typical hydrogen concentration in etched niobium is in the order
100~-~400~atomic~ppm hydrogen in the bulk material and 4~atomic
percent in a 50 nm surface layer \cite{roeth,bonin_roeth_91}. A high
temperature treatment at~800$^\circ$C is very effective in reducing
the hydrogen concentration to ~3~atomic ppm in the bulk and
$\leq$~1~atomic\% in the surface layer.

The hydride formation depends on the dwell time the cavity spends in
the dangerous region around 100~K. With a fast cooldown from 300~K to
4.2~K within 1~hour, hydride formation will be reduced to an acceptable
level. 

\begin{figure}[htbp]
\begin{center}
      \includegraphics[angle=-90,width=7cm]{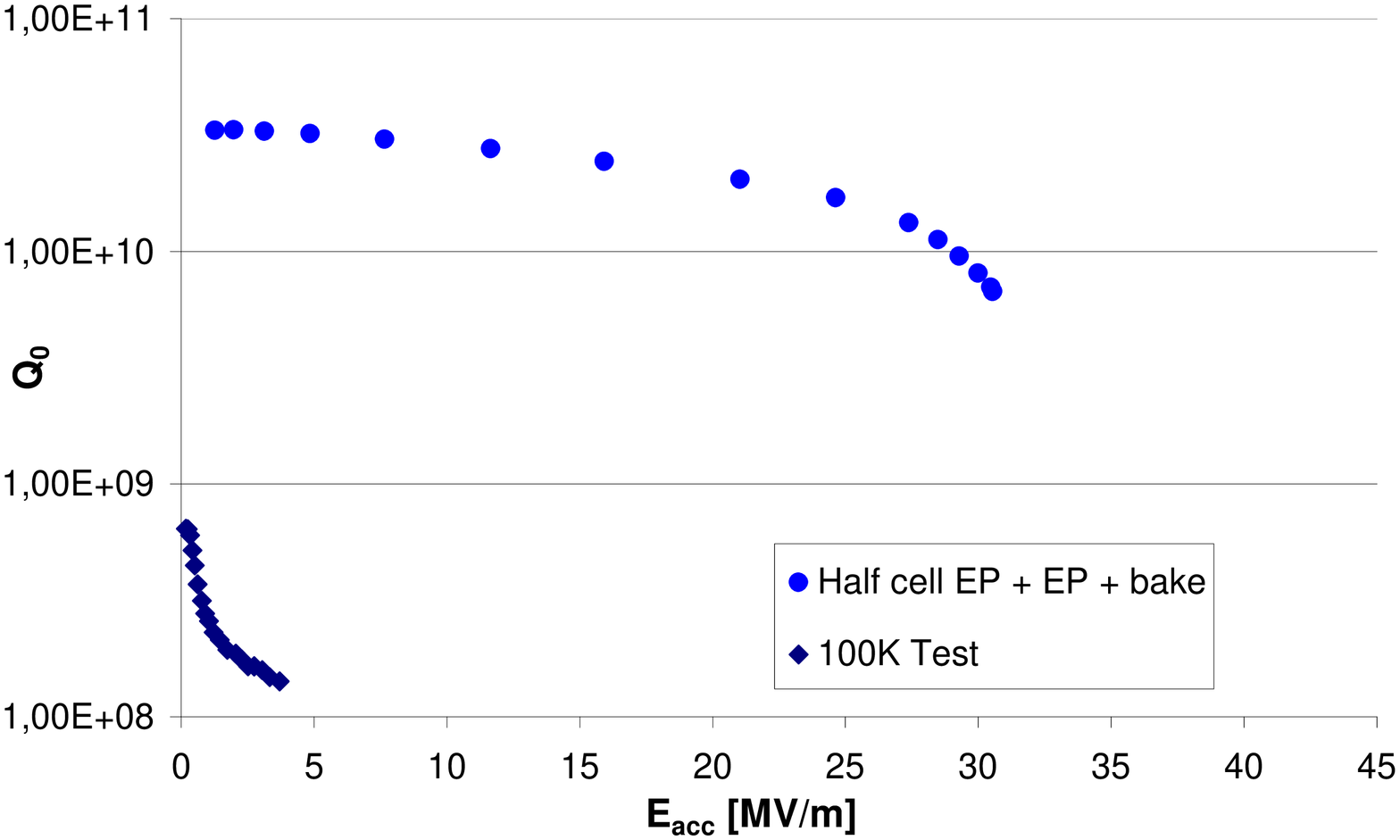} \newline 
      a)

      \includegraphics[angle=-90,width=7cm]{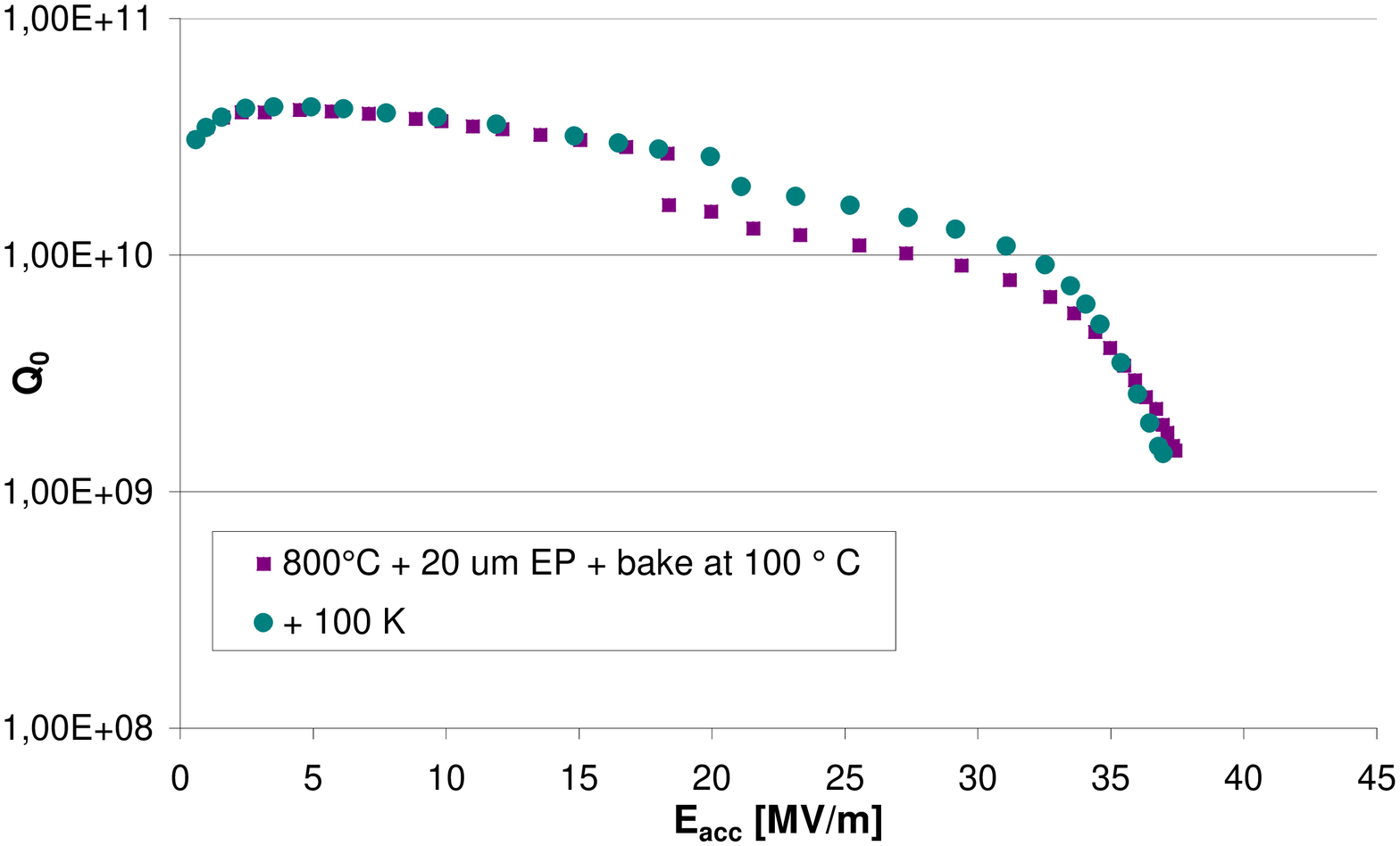} \newline 
      b)
  \caption{\label{fig:q_disease} Demonstration of the 
 ``Q disease'' i.e. the strong degradation due to the formation of
 niobium hydrides. a) The excitation curve of an electropolished and
 baked cavity before and after a two-day exposure to a 
 temperature of 100~K. b) The same cavity after an 800$^\circ$C
 furnace treatment and a short EP of 20~$\mu$m. The quality factor
 degradation  does not reappear  when the cavity is exposed to
 100~K. Test temperature 1.6~K. } 
\end{center}
\end{figure}

The cavities studied in this work have been made from electropolished
half-cells. In the half-cell EP system the niobium is not protected
from the hydrogen gas by a membrane. Therefore the danger of quality
factor degradation was to be expected. A dedicated experiment was
carried out to this end, the results are shown in figure
\ref{fig:q_disease}. The electropolished and baked cavity was tested
after a fast cooldown and showed the usual good performance.  Then the
cavity was exposed to a temperature of 100~K for 2 days.  In the new
test the quality factor was very low and showed the typical strong
field dependence described in \cite{roeth}.  In the next step the
cavity was heated in a vacuum furnace to 800$^\circ$C to remove the
hydrogen from the bulk material. A short electropolishing of about
20~$\mu$m was done to clean the surface from dirt which might have
been introduced during  the furnace treatment.  Again the cavity
was tested after fast cooldown and showed the initial good
performance.  The exposure of the cavity to the dangerous
temperature of 100~K for 2~days was repeated. This time, however, no
$Q$ degradation was observed, indicating  that the formation of
hydrides was strongly suppressed by the furnace treatment. This is an
important and encouraging result which implies that no hydrogen will
be introduced into the niobium during a short EP if precautions are
taken (PTFE-cloth around
the cathode) to keep the gas away from the niobium surface.

\begin{figure}[htbp]
  \begin{center}
  \includegraphics[angle=0,width=7cm]{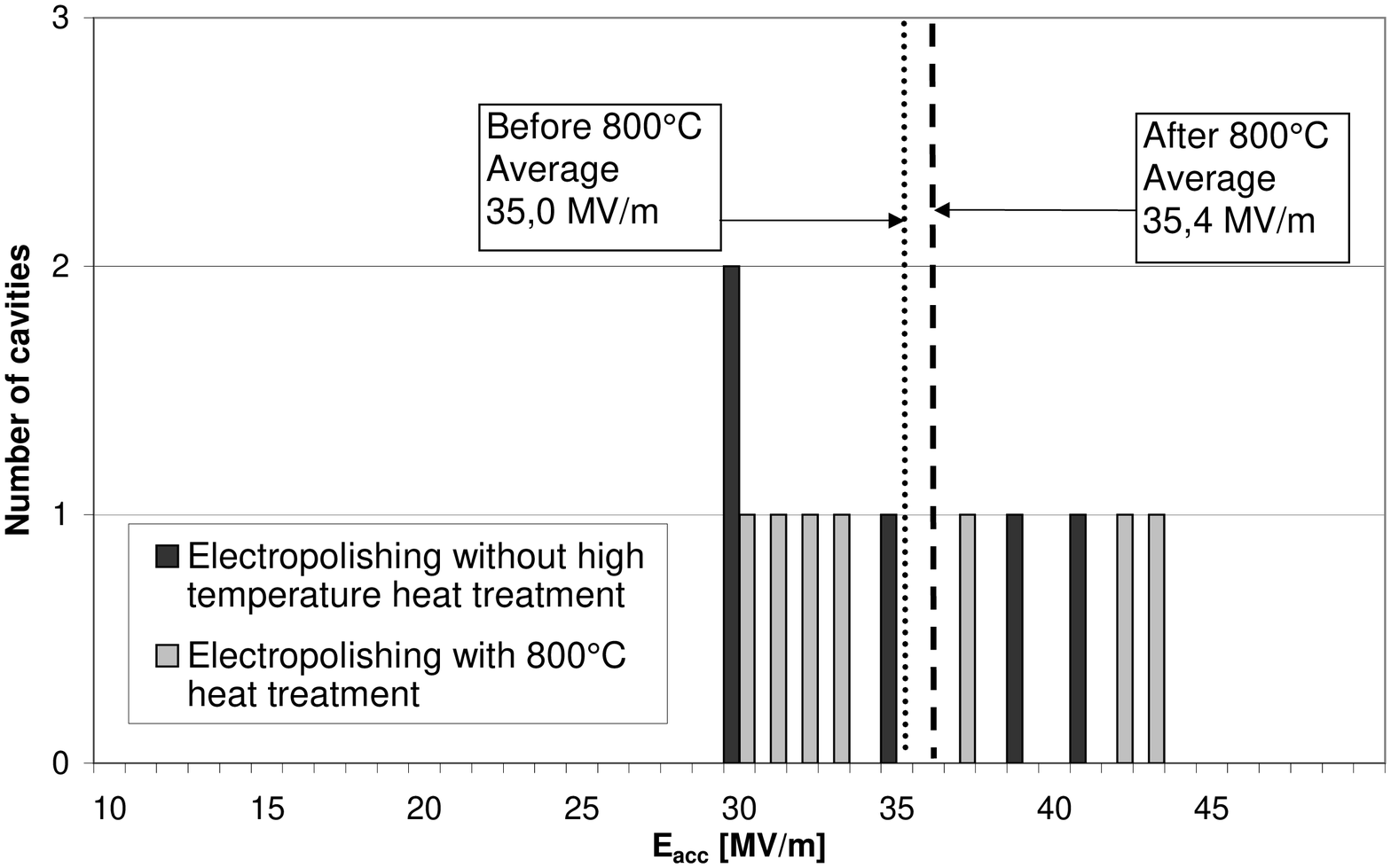}
    \caption{Distribution of accelerating
      gradients before  and after  the furnace treatment at 800$^\circ$C.
      One cavity with a weld defect has been omitted from this graph.}
    \label{fig:ep_800_stat}
  \end{center}
\end{figure}

Seven cavities received the furnace treatment at 800$^\circ$C after
electropolishing. Figure \ref{fig:ep_800_stat} shows the maximum
gradients before and after the 800$^\circ$C treatment. The average
gradient rises from 35~MV/m to 35.4~MV/m.

Finally, one electropolished cavity was subjected to the $1400^\circ$C
furnace treatment with titanium getter which is part of the standard
preparation of the TTF cavities (see \cite{tesla_cavities}). The idea
was to check whether the increased thermal conductivity would have a
similar beneficial effect on EP cavities as on BCP cavities. The
residual resistivity ratio RRR increased from ~300 to 500. The cavity
reached the same maximum field of 35~MV/m as before the 1400$^\circ$C
treatment, but at a slightly higher $Q_0$.
 More tests are needed to decide whether the $1400^\circ$C
furnace treatment would constitute a significant improvement of the
electropolished multi-cell cavities of the TESLA-800 collider.

\subsection{Comparison of etching and electropolishing}
\label{sec:comp_BCP_EP}
The test results on etched single-cell cavities are consistent with
the performance achieved in the TESLA nine-cell cavities.  The average
gradient achieved in the one-cell cavities without any heat treatment
is 24~MV/m.  If one evaluates the single-cell performance of those
etched nine-cell cavities which were tested after the 800$^\circ$C
heat treatment an average gradient of 23.5~MV/m is obtained.  The
1400$^\circ$C heat treatment which is routinely applied to the
BCP-treated TTF cavities raises the average gradient to more than 25~MV/m
\cite{tesla_cavities}.
\begin{figure}[!tbp]
  \begin{center}
  \includegraphics[angle=90,width=7cm]{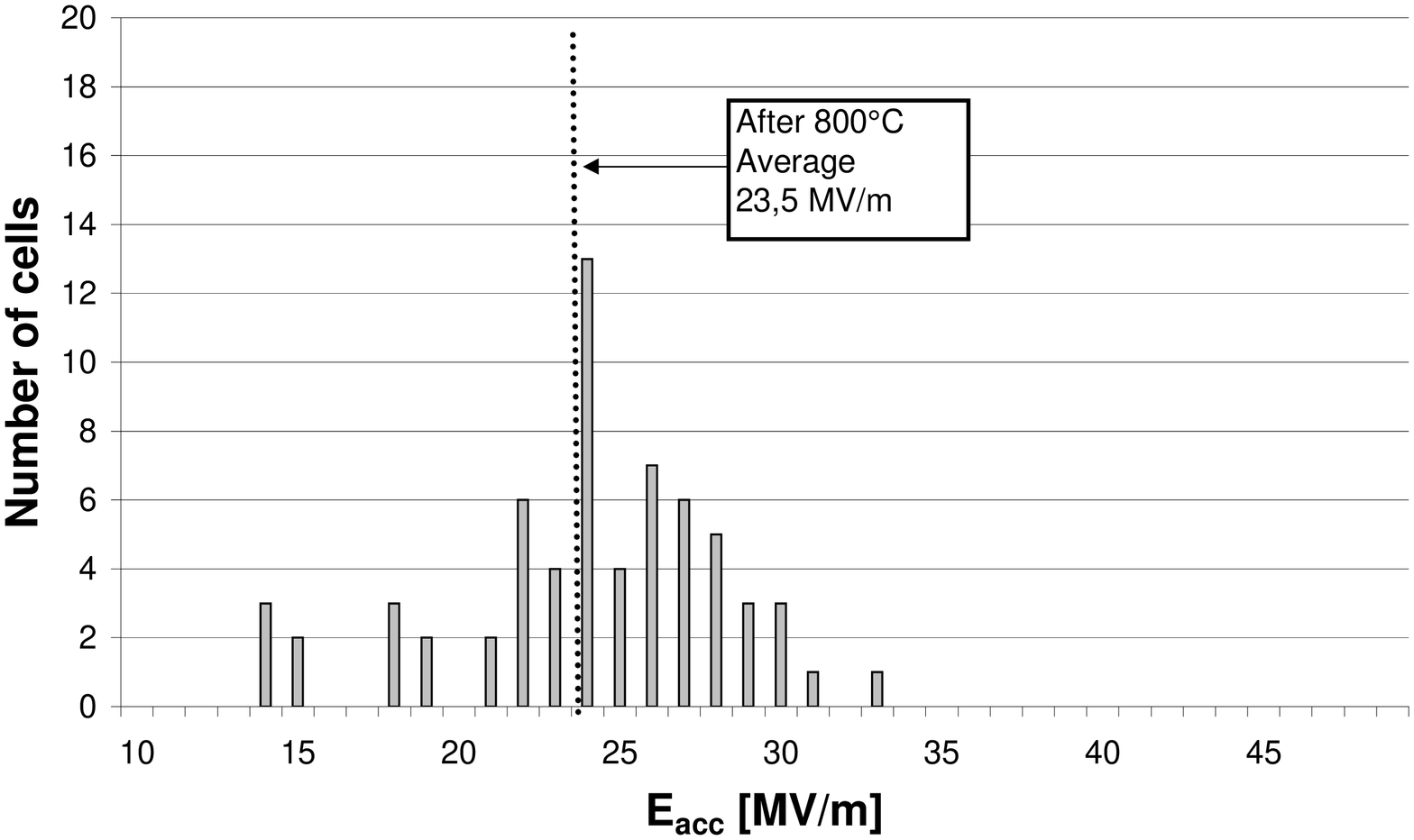}  \newline 
        a)

  \includegraphics[angle=-90,width=7cm]{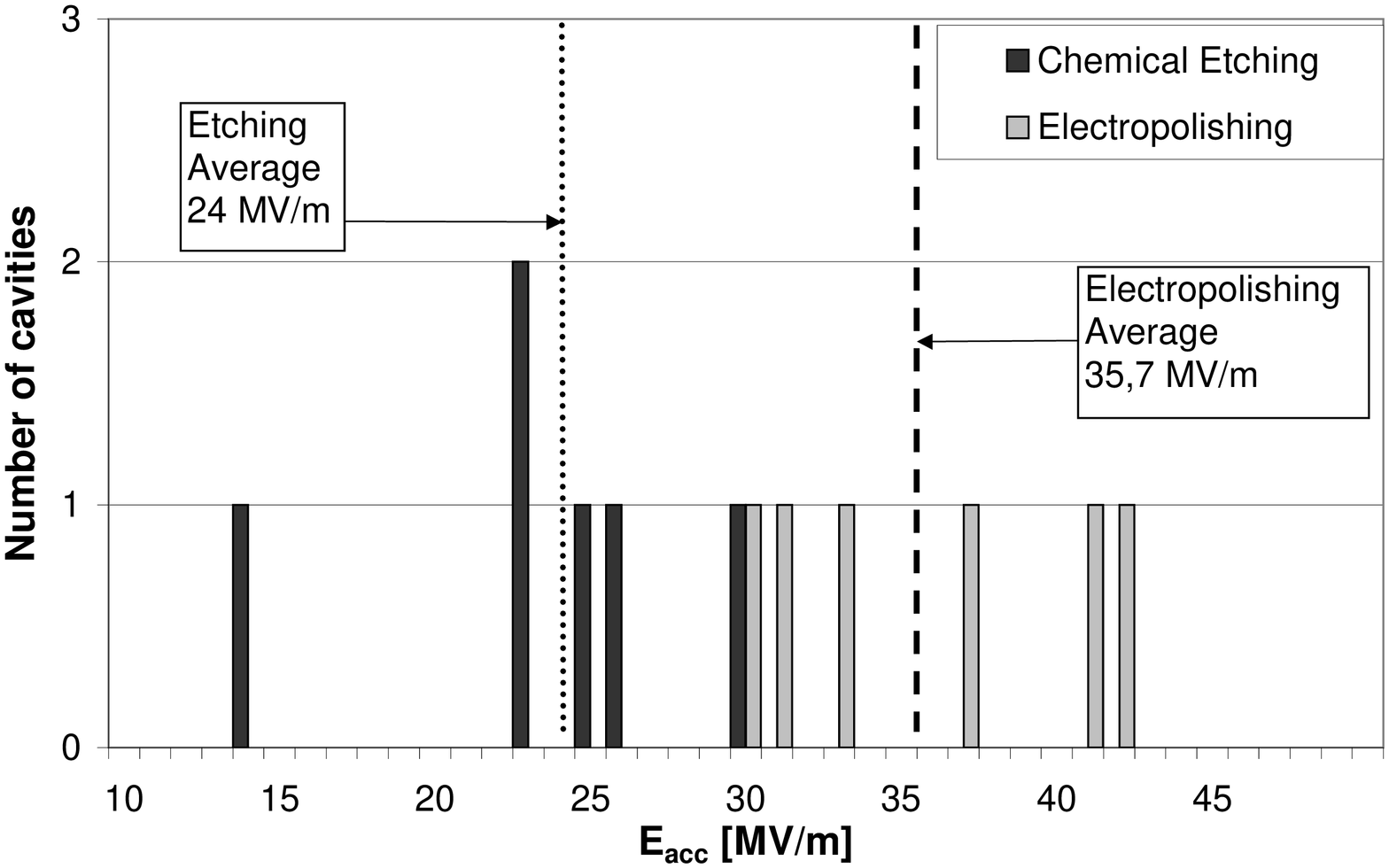}  \newline 
        b)
    \caption{a) Single-cell performance of the
      nine-cell TESLA cavities after the 800$^\circ$C heat treatment.
      The average gradient is 23.5~MV/m. b) Distribution
      of the maximum accelerating gradients of etched and
      electropolished single-cell cavities. For etched cavities the
      average gradient is 24~MV/m, for electropolished cavities
      35.7~MV/m. }
    \label{fig:bcp_ep_stat}
  \end{center}
\end{figure}
\begin{figure}[htbp]
  \begin{center}
  \includegraphics[angle=-90,width=7cm]{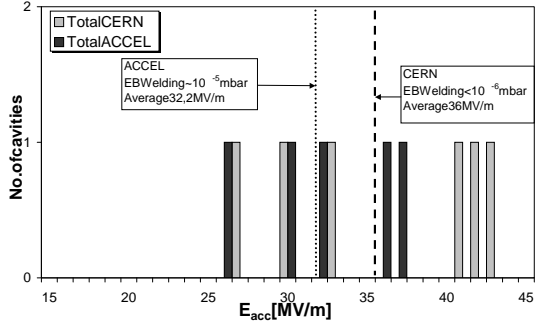}
    \caption{Comparison of
      accelerating gradients of cavities welded under standard
      vacuum conditions ($\approx 10^{-5}$~mbar) by the company ACCEL
      and under improved vacuum conditions ($<~10^{-6}$~mbar) in the
      electron beam welding machine at CERN. The data for etched and
      electropolished cavities are shown. The best test on each cavity
      is taken. One cavity of the CERN batch had a macroscopic defect
      on a weld seam, which was found during visual inspection. This
      cavity has been omitted from the figure.} 
    \label{fig:cern_accel}
  \end{center}
\end{figure}

A summary of the performance after etching and electropolishing is
given in figure \ref{fig:bcp_ep_stat}. The average gradient of the
electropolished cavities is 35.7~MV/m. This is convincing evidence
that electropolishing leads to substantially higher accelerating
fields.
 One reason is certainly that magnetic  field enhancements at grain boundaries
 are avoided. Moreover,  the smoother EP surface contains fewer defects
and is easier to clean from chemical residues. 

\subsection{Influence of the electron-beam welding on the cavity
  performance}
\label{sec:ebw_ep}
The TESLA cavities are made from deep-drawn niobium half-cells which
are joined by electron-beam (EB) welding.  The quality of the EB weld
seam at the equator is crucial for the maximum field achievable in the
cavities. The one-cell production was made in two different EB welding
machines. The first batch of 4 cavities was produced by an industrial
company with experience in niobium cavity welding (ACCEL Instruments,
Bergisch-Gladbach).  The second batch of 11 cavities was welded at
CERN with an EB welding machine equipped with a stainless steel vacuum
chamber conforming to ultra high vacuum standards. Whereas the vacuum
in the industrial production is of the order of a few times
$10^{-5}$~mbar, the pressure in the CERN system could be lowered down
to $3 \cdot 10^{-7}$~mbar by using a cryopump and a longer time for
pumping down. 

As the residual gas in the machine is partly gettered by the molten
niobium in the weld seam, the residual resistivity ratio 
RRR in this region can be degraded. For an initial RRR~$\approx$~300
the reduction can be of the order of 20\% for a
vacuum of $10^{-5}$ mbar 
\footnote{For TIG (Tungsten-Inert-Gas) welds the reduction can
be very large (RRR~$\approx$~400 is reduced to RRR~$\approx$~120)
\cite{singer_98}. In addition, defects can be introduced using this
welding technique.}. It can be expected that the better vacuum conditions
lead to an improved cavity performance.  The average gradient
of the CERN batch is in fact 3.8~MV/m higher as compared to the
cavities prepared under standard vacuum conditions (see figure
\ref{fig:cern_accel}) \footnote{All cavities were treated in a vacuum
  furnace at 800$^\circ$C to remove hydrogen from the niobium.}.

The CERN EB machine is equipped with a 60 kV electron gun
without the possibility to raster the electron beam in a rhombic
pattern across the weld which produces the smoothest underbead of the
weld seam \cite{tesla_cavity_tdr,padamsee_book}.  The results in
figure~\ref{fig:cern_accel} indicate that the electron-beam welding facility
should  combine  a state-of-the-art electron gun and a UHV vacuum
system. Such an electron-beam welding facility is presently being
commissioned at DESY.
\begin{figure}[!ht]
  \begin{center}
    \centering \includegraphics[angle=-90, width=7cm]{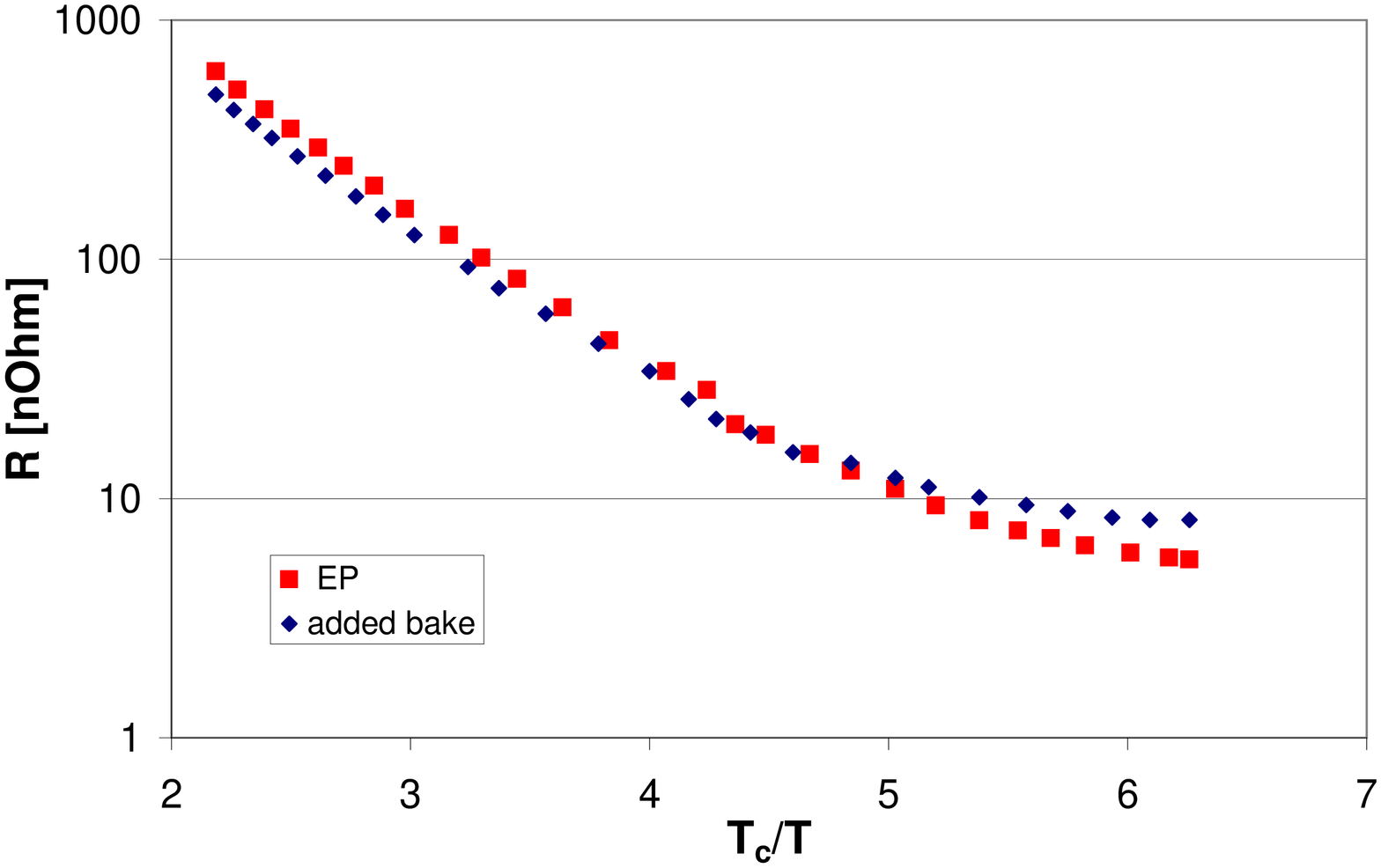} \newline
    a) 

    \centering \includegraphics[angle=-90, width=7cm]{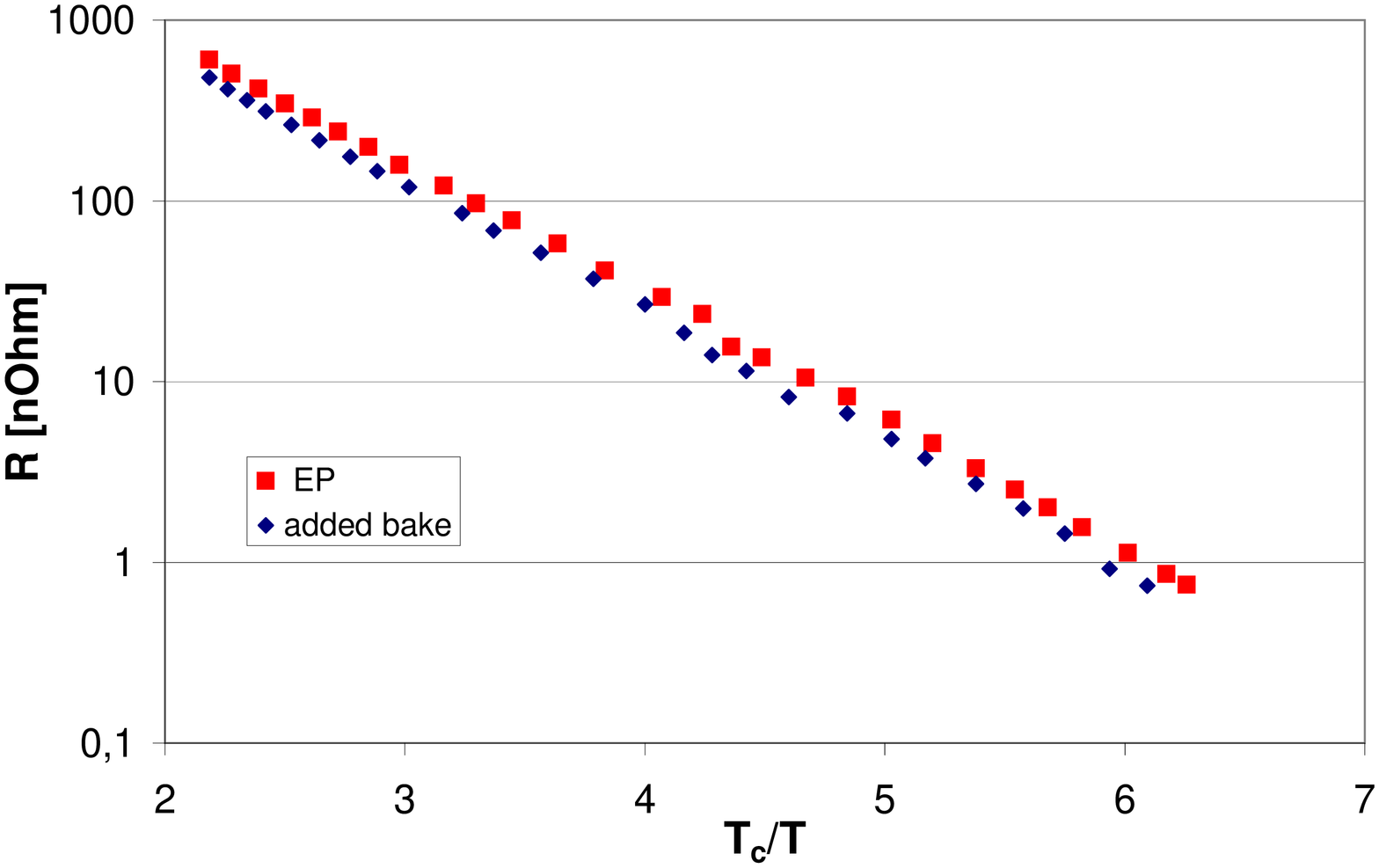} \newline
    b)
    \end{center}
\caption[Bakeout in a cryostat with helium atmosphere. ]{a)
      Temperature dependence of the surface  resistance before and after
     bakeout of an EP cavity. b) The BCS surface resistance. The  residual
    resistances of 4.7 n$\Omega$ for the unbaked cavity 
   and of 7.4 n$\Omega$ for the baked cavity have been subtracted.}
  \label{fig:rbcs_bake}
\end{figure}

\subsection{Influence of bakeout on the microwave surface resistance}
\label{sec:qd_bcs}
In figure \ref{fig:rbcs_bake} the microwave surface resistance before
and after the low-tem\-perature bakeout is plotted against the 
inverse temperature. According to the  BCS (Bardeen-Cooper-Schrieffer)
theory the surface resistance should drop exponentially when plotted
against $T_c/T$. For temperatures $T < \frac{T_c}{2}$ and
an energy of the microwave photons  $h f$ much less than the energy
gap $\Delta$, the surface
resistance can be approximated by:
\begin{equation}
  \label{eq:bcs_resistance}
  R_{BCS}(T,f) = A \frac{f^2}{T} \exp \left( -\frac{\Delta}{k_B T} \right)
\end{equation}
The factor $A$ depends on material parameters like the coherence
length $\xi$, the 
electron mean free path $\ell$, the Fermi velocity $v_F$  and the
London penetration depth $\lambda_L$.  The  BCS resistance of niobium
at 1.3~GHz is about 600~n$\Omega$ at 4.2~K and  10~n$\Omega$ at 2 K.

A refined expression, derived  from the two-fluid model of
superconductors, is
\cite{bonin_cas}:
\begin{equation}
  \label{eq:r_bcs_two_fluid}
  R_{BCS} (T,f) \propto  \sigma_{nc} \cdot \lambda_{eff}^3 \frac{f^2}{T} \exp \left( -\frac{\Delta}{k_B T} \right)
\end{equation}
where $\sigma_{nc}$ is the conductivity due to the normal-conducting
component and $\lambda_{eff}=\lambda_L \, \sqrt{1+ \xi/ \ell}$.
The BCS resistance is hence proportional to the mean free
path $\ell$ of the normal-conducting electrons and does not assume its
minimum for extremely pure niobium ($\ell$ very large) but rather for
somewhat ``dirty'' Nb with $\ell \approx \xi$.  If we make the assumption
that the bakeout creates a ``dirty'' surface layer with a shorter mean
free path than in the bulk niobium, then the residual resistance will
be larger and the BCS resistance will be lower than in the unbaked cavity. 
Figure \ref{fig:rbcs_bake} shows that this is indeed the case. 
The BCS part of the  surface resistance 
is reduced by a factor of $\approx 1.5$ due to the bakeout while the 
residual resistance  increases.
 Similar observations were
made by Kneisel \cite{kneisel_99}. 

It should be noted that the resistance data are derived from measurements
of the cavity quality factor at low field (a few MV/m). The peculiar
high-field behaviour of unbaked cavities, namely the strong $Q$
degradation, cannot be explained  in terms of the BCS resistance
because this quantity is independent  of the magnitude of
the electromagnetic field in the cavity.  On the contrary, a strongly
field-dependent resistance would be needed to account for the rapid
decrease of $Q_0$ towards large gradients. This effect is not yet
understood. 

\section{Conclusions and outlook}
\label{sec:conclusion-outlook}
Electropolished bulk niobium cavities offer accelerating fields of
more than 35 MV/m and are hence suitable for the upgrade of the TESLA
collider to 800 GeV.  The surface preparation by electropolishing has
a definite advantage in comparison with chemical etching: a much
smoother surface is obtained and the field enhancement at grain
boundaries is avoided. This magnetic field enhancement may be the
origin of a local premature breakdown of superconductivity in etched
cavities. The physical and physico-chemical principles of the bakeout
effect are not well  understood yet and require further research.
Meanwhile, the first multi-cell cavities have been electropolished in
a collaboration between KEK and DESY.  The results will be published
in a separate paper.

\section{Acknowledgments}
The authors would like to express their gratitude to O.~Aberle,
S.~Ehmele, S.~Forel, A.~Lava, A.~Insomby, R.~Guerin,
M.~Kubly, the DESY group MKS3, K.~Twarowski, J.-P.~Poupeau, Y.~Gasser for
their invaluable help in measurements, cavity fabrication, chemical
treatments and handling of cavities.

We are particularly grateful to K.~Saito and E.~Kako for their advice
and information on electropolishing.
We would like to thank E.~Haebel for the discussions on RF
superconductivity and help with the measurement system at CERN.
We would like to thank P.~Kneisel for stimulating discussions
on the  bakeout effect.

%
%

\bibliographystyle{unsrt}
\bibliography{LL_bibliography}

\end{document}